\documentclass[conference]{IEEEtran}
\usepackage[left=0.54in, right=0.54in, top=0.6in, bottom=0.6in,centering]{geometry}
\usepackage{bbm}
\usepackage{color}
\usepackage{graphicx}
\usepackage{adjustbox}
\usepackage[dvipsnames]{xcolor}
\usepackage{cite,algorithm,algorithmic,amsmath,amssymb,amsthm,empheq}
\usepackage{graphicx,graphics}
\usepackage{subfigure,amsfonts,balance}
\usepackage{epstopdf}
\usepackage{lipsum}

\DeclareMathOperator{\C}{\mathbb{C}}
\DeclareMathOperator{\F}{\pmb{F}}
\DeclareMathOperator{\FF}{\mathcal{F}}

\DeclareMathOperator{\K}{\mathcal{K}}

\DeclareMathOperator{\E}{\pmb{E}}

\DeclareMathOperator{\HH}{\pmb{H}}

\DeclareMathOperator{\RR}{\pmb{R}}

\DeclareMathOperator{\CN}{\mathcal{CN}}

\DeclareMathOperator{\aaa}{\pmb{a}}

\DeclareMathOperator{\MM}{\mathcal{M}}

\DeclareMathOperator{\PPP}{\mathcal{P}}

\DeclareMathOperator{\x}{\pmb{x}}
\DeclareMathOperator{\y}{\pmb{y}}
\DeclareMathOperator{\s}{\pmb{s}}
\DeclareMathOperator{\n}{\pmb{n}}

\DeclareMathOperator{\X}{\pmb{X}}

\DeclareMathOperator{\one}{\mathbbm{1}}

\DeclareMathOperator{\SIGMA}{\pmb{\Sigma}}

\DeclareMathOperator{\PHI}{\pmb{\Phi}}

\DeclareMathOperator{\PI}{\pmb{\Pi}}

\DeclareMathOperator{\XI}{\pmb{\Xi}}

\theoremstyle{remark}
\newtheorem{remark}{Remark}

\begin{document}
{
\title{\Huge Joint Optimization of User Association, Data Delivery Rate and Precoding for Cache-Enabled F-RANs}
}
\author{\IEEEauthorblockN{Tung T. Vu\IEEEauthorrefmark{1}, Duy T. Ngo\IEEEauthorrefmark{1}, Lawrence Ong\IEEEauthorrefmark{1}, Salman Durrani\IEEEauthorrefmark{2} and Richard H. Middleton\IEEEauthorrefmark{1}}
\IEEEauthorblockA{\small\IEEEauthorrefmark{1}School of Electrical Engineering and Computing, The University of Newcastle, Callaghan NSW 2308, Australia
\\ Email: thanhtung.vu@uon.edu.au, \{duy.ngo, lawrence.ong, richard.middleton\}@newcastle.edu.au}
\IEEEauthorblockA{\small\IEEEauthorrefmark{2}Research School of Engineering, The Australian National University, Canberra ACT 2601, Australia
\\ Email: salman.durrani@anu.edu.au}
}
\maketitle
\begin{abstract}

This paper considers the downlink of a cache-enabled fog radio access network (F-RAN) with limited fronthaul capacity, where user association (UA), data delivery rate (DDR) and signal precoding are jointly optimized.
We formulate a mixed-integer nonlinear programming problem in which the weighted difference of network throughput and total power consumption {is} maximized, subject to {the} predefined DDR requirements and {the} maximum transmit power at each eRRH.
To address this challenging problem, we first apply the $\ell_0$-norm approximation and $\ell_1$-norm minimization techniques to deal with the UA.
After this key step, we arrive at an approximated problem that only involves {the} joint optimization of DDR and precoding.
By using the alternating descent method, we further decompose this problem into a convex subproblem for DDR allocation and a nonconvex subproblem for precoding design.
While the former is globally solved by the interior-point method, the latter is solved by a specifically tailored successive convex quadratic programming method.
Finally, we propose an iterative algorithm for the original joint optimization that is guaranteed to converge.
Importantly, each iteration of the developed algorithm only involves solving simple convex problems.
Numerical examples demonstrate that the proposed design significantly improves both throughput and power performances, especially in practical F-RANs with limited fronthaul capacity.
Compared to the sole precoder design for a given cache placement, our joint design is shown to improve the throughput by $50\%$ while saving at least half of the total power consumption in the considered examples.
\end{abstract}
%
\IEEEpeerreviewmaketitle
\balance
\section{Introduction}
\label{sec:introd}
A fog radio access network (F-RAN) is recently proposed as an alternative to the cloud radio access network (C-RAN) to support mobile edge computing \cite{liu2017,Chiang2016,shi2015}.
Capable of exploiting the advantages of both local caching and centralized signal processing, this novel network architecture is expected to significantly improve both spectral and energy efficiencies of the fifth-generation (5G) of cellular systems \cite{Peng2016}.
In an F-RAN, traditional high-cost high-power base stations (BSs) are replaced by low-cost low-power enhanced remote radio heads (eRRHs).
Equipped with a finite-storage cache, each eRRH is connected to a central base band unit (BBU) via a fronthaul link.
If a user (UE) requests a file that is available at the local caches of its serving eRRHs, the file can be directly retrieved from the caches.
Otherwise, the file will be fetched from the BBU to the serving eRRHs before being transferred the UEs via radio access links.



In practical settings, the performance of an F-RAN is constrained by the capacity of its fronthaul links.
User association (UA) can help reduce the fronthaul traffic by assigning UEs to appropriate eRRHs and save power by putting the unassigned eRRHs into sleep mode \cite{liu2016}.
To minimize the fronthaul traffic and transmission power, the work of \cite{Tao2016} considers {the} joint design of multicast beamforming and BS clustering (which is essentially {a} UA problem) in cache-enabled C-RANs, where the BSs are assigned to the groups of UEs requesting the same file.
To maximize the throughput, {the} joint design of data delivery rate (DDR) and precoding has been studied in \cite{Park2016}.
However, since \cite{Park2016} assigns the UEs to the eRRHs heuristically, it may not exploit the full potential of UA to enhance the system performance.
\vspace{-0mm}

In this paper, we jointly design UA, DDR and signal precoding for the downlink of a cache-enabled F-RAN with limited fronthaul capacity.
We aim to maximize the weighted difference between the network throughput and the total power consumption, the latter of which consists of the operating power and the transmission power in both fronthaul and radio access links.
The formulated optimization problem is constrained on meeting the predefined DDR requirements, the limited fronthaul capacity and the maximum transmit power at each eRRH.
In this mixed-integer nonlinear program,
the strong coupling among the optimizing variables makes it even more challenging to be solved globally.

Here, we first express the UA variables as the functions of the $\ell_0$-norm of {the} precoding matrices.
We approximate this $\ell_0$-norm with its weighted $\ell_1$-norm \cite{Candes2008} and {update} the weight factor iteratively.
We then obtain an approximated problem in the variables of DDRs and precoding matrices only.
Next, we apply the alternating descent method in \cite{Bertsekas1999} {to decompose the approximated problem into a convex subproblem {for the DDR allocation} and a nonconvex subproblem for {the precoder design}.
The former is readily solved by a convex solver, whereas the latter is dealt with by the successive convex quadratic programming framework of \cite{Tam2016}.
Finally, we propose an iterative algorithm that is proved to converge once initialized from a feasible point.
Each iteration of the algorithm corresponds to solving simple convex programs.

In the numerical examples with practical parameter settings, the proposed design demonstrates its capability in substantially improving both throughput and power {performances}.
Compared to solely designing the precoders for a given cache placement, the developed joint design offers $50\%$ throughput gain while consuming only $50\%$ of the total power otherwise required.
The performance enhancement is particularly pronounced in cases with limited-capacity fronthaul links.

\emph{Notation:}
Boldfaced symbols are used for vectors and capitalized boldfaced symbols for matrices. $\pmb{X}^H$ is the conjugate transposition of a matrix $\pmb{X}$. $\pmb{I}$ and $\pmb{0}$ are the identity and zero matrix with the appropriate dimensions respectively. $||.||_0$ denotes the $\ell_0$-norm. $\langle\X\rangle$ means the trace of a matrix $\X$.

\section{System Model and Problem Formulation}
\label{sec:systemmodel}
Consider the general F-RAN model {\cite{Park2016}} illustrated in Fig.~\ref{fig:FRANModel}, where there are $K_U$ UEs capable of establishing wireless connections with $K_R$ RRHs. Each UE $k\in\K_U\triangleq\{1,\dots,K_U\}$ is equipped with $N_u$ antennas, whereas each eRRH $i \in \K_R\triangleq\{1,\dots,K_R\}$ with $N_r$ antennas. The eRRH $i\in \K_R$ connects to the baseband unit (BBU) in the core network via a fronthaul link of capacity $C_i>0$ (b/s). \vspace{-2mm}
\begin{figure}[t!]
\centering
\includegraphics[width=0.35\textwidth]{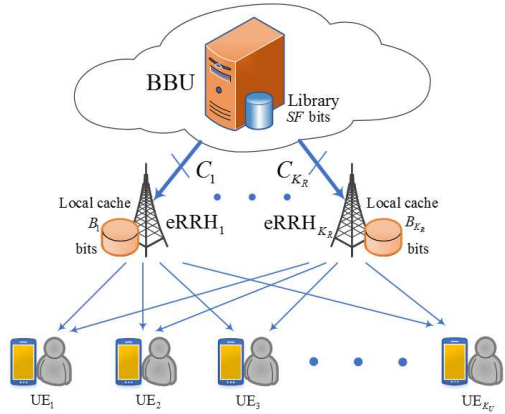}
\caption{Illustration of a general F-RAN}
\label{fig:FRANModel}
\end{figure}

\subsection{Data Request and Pre-fetching}
First, in the data request phase each UE $k\in\K_U$ requests a {random} file $f_k$ from the library $\FF\triangleq\{1,...,F\}$ stored in the BBU. Without loss of generality, assume that all files in the library are of the same size $S$ bits; hence, the total size of the file library is $SF$ bits. Denote by $\FF^{*}\triangleq\bigcup_{k\in\K_U}\{f_k\}$ the set of all the files requested by all $K_U$ users. To bring data contents closer to the UEs, each eRRH $i\in\K_R$ is equipped with a local cache that can store $B_i>0$ bits.
%

It is practical to assume that the eRRHs have limited storage capacity and therefore only a subset of the file library is cached at each eRRH. In this paper, we adopt the fractional cache distinct strategy \cite{Park2016}. Each file in the library is split into $M$ subfiles of equal size $\bar{S}=S/M$ bits. Each eRRH $i\in\K_R$ then randomly selects the set $\MM^i$ of $\lfloor \frac{B_i}{\bar{S}}\rfloor$ subfiles from the BBU file library to store in its local cache. {The cache state information, which shows whether the subfile $(f_k,m)$ is cached at eRRH $i$ or not}, $\ f_k\in\FF^{*}, \ m\in \MM \triangleq\{1,\dots,M\}$  is summarized as \vspace{-4mm}

{
\small
\begin{align}\label{CDset}
c_{f_k,m}^i &\triangleq
\left\{
\begin{array}{ll}
  1, & \text{if $(f_k,m)\in\MM^i$} \\
  0, & \text{otherwise}
\end{array}.
\right.
\end{align}
}
\vspace{-4mm}
%
%

With data pre-fetching, the files requested by the UEs can now be retrieved directly from the local cache of the serving eRRHs instead of from the BBU.
If the subfile $(f_k,m)$ requested by UE $k\in\K_U$ is not available at eRRH $i$'s local cache, $(f_k,m)$ will be fetched from the BBU via the fronthaul link.

In each data-request duration, the requested files and local caches are known.
The cache state information $c_{f_k,m}^i \ (f_k\in\FF^{*}, \ m\in \MM, \ i\in\K_R)$ is thus available at the BBU.
In the following, we focus on the data delivery phase from the BBU and/or the eRRHs to the UEs, where each optimization and transmission block (OTB) interval is assumed to be shorter than the data-request duration.
We do not optimize across multiple OTBs but instead only in one OTB.

\vspace{-3mm}

\subsection{Data Delivery}\label{ssec:data_deliver}
In the considered F-RAN, the central BBU allows each eRRH to serve multiple UEs and each UE to be served by multiple eRRHs. We model the associations of eRRHs and UEs by the following binary variables: \vspace{-5mm}

{
\small
\begin{align}\label{aselect}
a_{k,i} \triangleq
\left\{
\begin{array}{ll}
  1, & \text{if eRRH $i$ serves UE $k$} \\
  0, & \text{otherwise}
\end{array}
\right.
\end{align}
}
\vspace{-4mm}

Let $R_{f_k,m} \leq \bar{S}$ be the data delivery rate (DDR) of subfile $(f_k,m)$. Here, we allow $R_{f_k,m}$ to be transferred to UE $k\in\K_U$ in the considered OTB interval and leave $\bar{S}-R_{f_k,m}$ bits for the next OTB.

First, let us consider the data transmission from the BBU to the eRRHs.
At this step, the BBU needs to decide the set of associated eRRHs to which each missing requested subfile $(f_k,m)$ is transferred via the fronthaul links.
There are trade-offs between transferring all missing requested subfiles and only transferring selected subfiles to the associated eRRHs.
While the former requires more data to be transferred via fronthaul links, it can give more throughput gain as a result of coherence combining.
In this paper, we adopt the former approach. The total data rate on the fronthaul link of eRRH $i\in\K_R$ can therefore be expressed as: \vspace{-5mm}

{
\small
\begin{align}\label{FH:cons}
R_i^{FH}\triangleq\sum_{f_k\in\FF^{*}}a_{k,i}\sum_{m\in\MM}(1-c_{f_k,m}^i)R_{f_k,m}.
\end{align}
}\vspace{-3mm}

Next, we consider the data transmission from the eRRHs to the UEs. Let $\s_{f_k,m}\in\C^{d\times 1}$ denote the encoded baseband signal of subfile $(f_k,m)$ with $d$ data streams. Assume that $\s_{f_k,m}\sim\CN(\pmb{0},\pmb{I})$. The eRRH $i$ precodes $\s_{f_k,m}$ by a matrix $\F_{f_k,m}^i\in\C^{N_{r}\times d}$ to obtain the transmitted signal $\x_i=\sum_{f_k\in\FF^{*}}\sum_{m\in\MM}\F_{f_k,m}^i\s_{f_k,m}$.
Denote by $\HH_{k,i}\in\C^{N_{u}\times N_{r}}$ the flat-fading channel matrix from eRRH $i$ to UE $k$ and by $\HH_k\triangleq[\HH_{k,1},...,\HH_{k,K_R}]\in\C^{N_{u}\times N_R}$ the channel matrix from all eRRHs to UE $k$, where $N_R \triangleq K_RN_{r}$. Assume that channel states $\HH_{k,i}, \ k\in\K_U, \ i\in\K_R$ remain unchanged during each OTB and are made available to the BBU and eRRHs.

Define 
$\bar{\F}_{f_k,m}\triangleq\left[(\F_{f_k,m}^1)^H,
  (\F_{f_k,m}^2)^H,
  \dots
  (\F_{f_k,m}^{K_R})^H\right]^H \in\C^{N_R\times d}$.
Note that since $\F_{f_k,m}^i = \pmb{0}$ when eRRH $i$ does not serve UE $k$. Then, the received signal $\y_k\in\C^{N_{u}\times 1}$ at UE $k$ for the requested file $f_k$ is: \vspace{-5mm}

{
\small
\begin{align}\label{yk:full}
\nonumber
\y_k \triangleq
&\HH_k\bar{\F}_{f_k,m}\s_{f_k,m}+\sum_{q\in\MM\setminus \{m\}}\HH_k\bar{\F}_{f_k,q}\s_{f_k,q}
\\
&+\sum_{f_\ell\in\FF^{*}\setminus \{f_k\}}\sum_{m\in\MM}\HH_k\bar{\F}_{f_\ell,m}\s_{f_\ell,m}
+ \n_k,
\end{align}
}where $\n_k\sim\CN(\pmb{0},\SIGMA_k)$ is the additive noise term.

{To deal with the interference expressed in the second and third terms on the right-hand side of \eqref{yk:full}}, we assume each UE $k$ performs the successive interference cancellation (SIC) decoding for the subfiles with the order $\s_{f_k,1}\rightarrow...\rightarrow\s_{f_k,M}$.
After applying the SIC scheme, {the achievable data rate $R_{f_k,m}$ can be bounded as \cite{Park2016,Park2016SPL}}: \vspace{-5mm}

{
\small
\begin{align}\label{eq:R_ub}
R_{f_k,m}
\leq g_{f_k,m}(\bar{\F})
\triangleq
\log
\left|
\pmb{I}_{N_{U,k}}+\PI_{f_k,m}\PI_{f_k,m}^H\XI_{f_k,m}^{-1}
\right|,
\end{align}
}\vspace{-3mm}

\noindent
where $\bar{\F}\triangleq\{\bar{\F}_{f_\ell,m}\}_{f_\ell\in\FF^{*},m\in\MM}$,
$\PI_{f_k,m}\triangleq\HH_k\bar{\F}_{f_k,m}$,
and \vspace{-0mm}
{
\small
\begin{align}\label{XI}
\nonumber
\XI_{f_k,m}\triangleq
&\sum_{q=m+1}^M\HH_k\bar{\F}_{f_k,q}\bar{\F}_{f_k,q}^H\HH_k^H +
\\
&\sum_{f_\ell\in\FF^{*}\setminus \{f_k\}}\sum_{q\in\MM}\HH_k\bar{\F}_{f_\ell,q}\bar{\F}_{f_\ell,q}^H\HH_k^H + \SIGMA_k.
\end{align}
}\vspace{-3mm}
%

\noindent
The network throughput is then defined as the following sum rate:\vspace{-4mm}

{
\small
\begin{align}\label{eq:sum_rate}
R_\text{sum}&\triangleq\sum_{f_k\in\FF^{*}}\sum_{m\in\MM}R_{f_k,m}.
\end{align}
}\vspace{-4mm}


\subsection{Power Consumption Model}
To see the effect of the joint design on the power performance, we adopt a practical power consumption model in \cite{Dai2016}.
Specifically, the per-OTB power consumption by eRRH $i\in\K_R$ is modeled as:\vspace{-5mm}

{
\small
\begin{align}\label{eRRH:Power}
P_i^{eRRH} \triangleq
\left\{
\begin{array}{ll}
  \beta_iP_i^{tx}+P_{i,a}, & \text{if $0<P_i^{tx}\leq P_i$} \\
  P_{i,s}, & \text{if $P_{i}^{tx} = 0$}
\end{array}
\right.,
\end{align}
}\vspace{-4mm}

\noindent
where constant $\beta_i>0, \ i\in\K_R$ reflects the power amplifier efficiency, feeder loss and other loss factors due to power supply and cooling for eRRH $i$ \cite{Dai2016};
$P_i^{tx}$ is the transmit power required to deliver all requested files from eRRH $i$ as \vspace{-4mm}

{
\small
\begin{align}\label{Pi:Tx}
P_i^{tx}&\triangleq\sum_{f_k\in\FF^{*}}\sum_{m\in\MM}\langle\bar{\E}_{i}^H\bar{\F}_{f_k,m}\bar{\F}_{f_k,m}^H\bar{\E}_{i}\rangle,
\end{align}
}in which  $\bar{\E}_i\in\C^{N_R\times N_{r}}$ is zero everywhere except an identity matrix of size $N_r$ from row $(i-1) N_r+1$ to row $i N_r$; $P_{i,a}$ is the power required to support eRRH $i$ in the active mode; and $P_{i,s} < P_{i,a}$ is the power consumption in the sleep mode.

The fronthaul link from the BBU to eRRH $i\in\K_R$ is modeled as a set of communication channels with a total capacity $C_i$ and total power dissipation $P_{i,\max}^{FH}$. Its power consumption is given by \vspace{-5mm}

{
\small
\begin{align}\label{FH:Power}
P_i^{FH} &\triangleq \frac{R_i^{FH}}{C_i}P_{i,\max}^{FH}=\alpha_iR_i^{FH},
\end{align}
}where $\alpha_i\triangleq{P_{i,\max}^{FH}}/{C_i}$ and $R_i^{FH}$ is defined in \eqref{FH:cons}.

From \eqref{eRRH:Power} and \eqref{FH:Power}, the total network power consumption is: \vspace{-4mm}

{
\small
\begin{align}\label{total:power}
\nonumber
P_\text{total}
&\triangleq \sum_{i\in\K_R}(P_i^{eRRH}+P_i^{FH})
\\
&
= \sum_{i\in\K_R}\big(\beta_iP_i^{tx}+ \one_{\{P_i^{tx}\}}P_{i,\Delta}+\alpha_iR_i^{FH}\big) + P_{s},
\end{align}
}
where $P_{i,\Delta} \triangleq P_{i,a}-P_{i,s}$, $P_{s}\triangleq\sum_{i\in\K_R}P_{i,s}$ and \vspace{-4mm}

{
\small
\begin{align}\label{indicator}
\one_{\{P_i^{tx}\}}&\triangleq
\left\{
\begin{array}{cc}
  1, & \text{if $P_i^{tx}>0$} \\
  0, & \text{otherwise}
\end{array}
\right..
\end{align}
} \vspace{-4mm}

\subsection{Problem Formulation}
In this paper, we aim to jointly design the UA, DDR and precoding in order to improve the network sum rate in \eqref{eq:sum_rate} as well as to reduce the total power consumption in \eqref{total:power}.
Let us define $\aaa\triangleq\{a_{k,i}\}_{k\in\K_U,i\in\K_R}$ and $\RR\triangleq\{R_{f_k,m}\}_{f_k\in\FF^{*},m\in\MM}$. For a given cache state information $\{c_{f_k,m}^i\}_{i\in\K_R, f_k\in\FF^{*},m\in\MM}$, the design problem is formulated as:

\vspace{-3mm}
{
\small
\begin{subequations}\label{mainP:GEE}
\begin{align}
\underset{{\aaa}, \RR, \bar{\F}}{\max} \,\,
&\PPP_1(\RR,\bar{\F})\triangleq R_\text{sum}-\eta P_\text{total}
\\
\mathrm{s.t.}\,\,
&R_\text{QoS}\leq R_{f_k,m}\leq \bar{S},\forall f_k\in\FF^{*},m\in\MM,\label{eq:R_qos}
\\
&\sum_{f_k\in\FF^{*}}a_{k,i}\sum_{m\in\MM}(1-c_{f_k,m}^i)R_{f_k,m}\leq C_i,\forall i\in\K_R\label{eq:C_FH}
\\
&R_{f_k,m}\leq g_{f_k,m}(\bar{\F}), \forall f_k\in\FF^{*},m\in\MM,,\label{eq:R_ubound}
\\
&\sum_{f_k\in\FF^{*}}\sum_{m\in\MM}\langle\bar{\E}_{i}^H\bar{\F}_{f_k,m}\bar{\F}_{f_k,m}^H\bar{\E}_{i}\rangle
\leq P_i,\forall i\in\K_R\label{eq:Pi_max}
\\
&\text{and} \ \eqref{aselect}, \ \forall k\in\K_U, \ i\in\K_R.\label{eq:a_ki}
\end{align}
\end{subequations}
}\vspace{-3mm}

\noindent
Here, the weight $\eta>0$ specifies the relative importance between the sum rate and the total power. Constraint \eqref{eq:R_qos} imposes the minimum rate $R_\text{QoS}\geq 0$ and maximum rate $\bar{S}$ for each subfile [see the second paragraph of Sec.~\ref{ssec:data_deliver}]. Constraint \eqref{eq:C_FH} expresses the bottleneck at fronthaul link $i\in\K_R$ with the limited backhaul capacity $C_i\geq0$ [see \eqref{FH:cons}]. Constraint \eqref{eq:R_ubound} is indeed \eqref{eq:R_ub}.
Finally, constraint \eqref{eq:Pi_max} requires that the total transmit power at each eRRH $i\in\K_R$ must not exceed the predefined budget $P_i\geq 0$. 
{While problem \eqref{mainP:GEE} is already a difficult nonconvex nonsmooth optimization problem, the strong coupling among the optimizing variables makes it even more challenging to be solved globally.


{
}

\section{Proposed Joint Optimization Algorithm
}
\label{sec:proposedalgorithm}

First, we will deal with the nonsmooth nature of \eqref{eq:a_ki} and \eqref{indicator}.
We begin by expressing \eqref{eq:a_ki} as: \vspace{-3mm}

{
\small
\begin{align}
\label{assign:cont:GEE}
&a_{k,i} =
\left\{
\begin{array}{ll}
  0, & \text{iff  $\bar{\E}_{i}^H\bar{\F}_{f_k,m} = \pmb{0}$},\forall m\in\MM \\
  1, & \text{otherwise}
\end{array}
\right.
\end{align}
}\vspace{-3mm}

\noindent
$\forall k\in\K_U, i\in\K_R$. To see this, note that if eRRH $i$ does not serve UE $k$, all corresponding precoders $\F_{f_k,m}^i=\bar{\E}_{i}^H\bar{\F}_{f_k,m}$ must be $\pmb{0}$ and then $a_{k,i}=0$. Otherwise, $a_{k,i}=1$ and there exists at least one of the corresponding precoders $\F_{f_k,m}^i=\bar{\E}_{i}^H\bar{\F}_{f_k,m}\neq\pmb{0}$. Therefore, without loss of optimality, $a_{k,i}$ can be further rewritten as: \vspace{-3mm}

{
\small
\begin{align}\label{assignvar:apprxm}
a_{k,i} = \big\|\sum_{m\in\MM}\langle\bar{\E}_{i}^H\bar{\F}_{f_k,m}\bar{\F}_{f_k,m}^H{\bar{\E}_{i}}\rangle\big\|_0.
\end{align}
}\vspace{-3mm}

\noindent
Similarly, \eqref{indicator} can also be replaced by: \vspace{-3mm}

{
\small
\begin{align}\label{indicator:apprxm}
\one_{\{P_i^{tx}\}}
=\big\|P_i^{tx}\big\|_0
=\Big\|\sum_{f_k\in\FF^{*}}\sum_{m\in\MM}\langle{\bar{\E}_{i}}^H\bar{\F}_{f_k,m}\bar{\F}_{f_k,m}^H{\bar{\E}_{i}}\rangle\Big\|_0.
\end{align}
}\vspace{-3mm}

\noindent
We respectively approximate the nonconvex $\ell_0$-norms \eqref{assignvar:apprxm} and \eqref{indicator:apprxm} by their reweighted $\ell_1$-norms as \cite{Dai2016} \vspace{-3mm}

{
\small
\begin{align}
a_{k,i}
&= \mu_{k,i}\sum_{m\in\MM}\langle{\bar{\E}_{i}}^H\bar{\F}_{f_k,m}\bar{\F}_{f_k,m}^H{\bar{\E}_{i}}\rangle,\label{assignvar:l1norm}\\
\one_{\{P_i^{tx}\}}
&=\theta_i\sum_{f_k\in\FF^{*}}\sum_{m\in\MM}\langle{\bar{\E}_{i}}^H\bar{\F}_{f_k,m}\bar{\F}_{f_k,m}^H{\bar{\E}_{i}}\rangle,\label{indicator:l1norm}
\end{align}
}\vspace{-3mm}

\noindent
where weights $\mu_{k,i}$ and $\theta_i$ are iteratively updated  according to \vspace{-3mm}

{
\small
\begin{align}
\label{mu:update}
\mu_{k,i}&=\frac{c_1}{\sum_{m\in\MM}\langle{\bar{\E}_{i}}^H\bar{\F}_{f_k,m}\bar{\F}_{f_k,m}^H{\bar{\E}_{i}}\rangle+\tau_1},
\\
\label{theta:update}
\theta_i&=\frac{c_2}{\sum_{f_k\in\FF^{*}}\sum_{m\in\MM}\langle{\bar{\E}_{i}}^H\bar{\F}_{f_k,m}\bar{\F}_{f_k,m}^H{\bar{\E}_{i}}\rangle+\tau_2}.
\end{align}
}\vspace{-3mm}

\noindent
Here, $c_1$ and $c_2$ are constant numbers whereas $\tau_1$ and $\tau_2$ are constant regularization factors. 
With \eqref{assignvar:l1norm}, variable $\aaa$ can now be expressed in terms of $\bar{\F}$ and thus be removed from \eqref{mainP:GEE}. With \eqref{indicator:l1norm}, we can rewrite $P_{\text{total}}$ as: \vspace{-3mm}

{
\small
\begin{align}\label{Ptotal:apprxm}
P_\text{total}
=& \sum_{i\in\K_R}\sum_{f_k\in\FF^{*}}\tau_{f_k}^i\sum_{m\in\MM}
\langle{\bar{\E}_{i}}^H\bar{\F}_{f_k,m}\bar{\F}_{f_k,m}^H{\bar{\E}_{i}}\rangle
 + P_{s}
\end{align}
}

\noindent
where $\tau_{f_k}^i\triangleq\upsilon_i+\alpha_i\vartheta_{f_k}^i$, $\vartheta_{f_k}^i\triangleq\mu_{k,i}\sum_{m\in\MM}(1-c_{f_k,m}^i)R_{f_k,m}$ and
$\upsilon_i\triangleq\beta_i+\theta_iP_{i,\Delta}$.

The above result allows us to transform \eqref{mainP:GEE} to the following approximated problem: \vspace{-5mm}

{
\small
\begin{subequations}\label{mainP:GEE:equiv}
\begin{align}
\underset{\RR,\bar{\F}}{\max} \,\,
&\PPP_2(\RR,\bar{\F})\triangleq R_\text{sum}-\eta P_\text{total}
\\
\mathrm{s.t.}\,\,
&R_{QoS} \leq R_{f_k,m}\leq \bar{S},\forall f_k\in\FF^{*},m\in\MM,
\\
&\sum_{f_k\in\FF^{*}}\sum_{m\in\MM}\vartheta_{f_k}^i\langle{\bar{\E}_{i}}^H\bar{\F}_{f_k,m}\bar{\F}_{f_k,m}^H{\bar{\E}_{i}}\rangle\leq C_i,\forall i\in\K_R
\\
&R_{f_k,m}\leq g_{f_k,m}(\bar{\F}),\forall f_k\in\FF^{*},m\in\MM,
\\
&\sum_{f_k\in\FF^{*}}\sum_{m\in\MM}\langle{\bar{\E}_{i}}^H\bar{\F}_{f_k,m}\bar{\F}_{f_k,m}^H{\bar{\E}_{i}}\rangle
\leq P_i,\forall i\in\K_R.
\end{align}
\end{subequations}
}\vspace{-2mm}

\noindent
Problem \eqref{mainP:GEE:equiv} is still difficult due to the strong coupling between $\RR$ and $\bar{\F}$. As such, we propose using an alternating method \cite{Bertsekas1999} that deals with one variable at a time and repeats until convergence. Specifically,
for a given $\bar{\F}$, we solve the following convex subproblem to update $\RR$: \vspace{-2mm}
{
\small
\begin{subequations}\label{mainP:GEE:equiv:updateR}
\begin{align}
\underset{\RR}{\max} \,\,\,\,\,\,
&\sum_{f_k\in\FF^{*}}\sum_{m\in\MM}R_{f_k,m}
-\eta\left(b+\sum_{i\in\K_R}\sum_{f_k\in\FF^{*}}\sum_{m\in\MM}q_{f_k,m}^iR_{f_k,m}\right)
\\
\mathrm{s.t.}\,\,\,\,\,\,
&R_{QoS}\leq R_{f_k,m}\leq \bar{S},\forall f_k\in\FF^{*},m\in\MM,
\\
&\sum_{f_k\in\FF^{*}}q_{f_k}^i\sum_{m\in\MM}R_{f_k,m}\leq C_i,\forall i\in\K_R,
\\
&R_{f_k,m}\leq g_{f_k,m}(\bar{\F}), \forall f_k\in\FF^{*},m\in\MM,
\end{align}
\end{subequations}
}\vspace{-3mm}

\noindent
where
{
\small
$q_{f_k}^i\triangleq\alpha_i\mu_{k,i}\sum_{m\in\MM}\langle{\bar{\E}_{i}}^H\bar{\F}_{f_k,m}\bar{\F}_{f_k,m}^H{\bar{\E}_{i}}\rangle(1-c_{f_k,m}^i)$
} and
{
\small
$b\triangleq P_{s}+\sum_{i\in\K_R}\sum_{f_k\in\FF^{*}}\sum_{m\in\MM}\upsilon_i\langle{\bar{\E}_{i}}^H\bar{\F}_{f_k,m}\bar{\F}_{f_k,m}^H{\bar{\E}_{i}}\rangle$.
}

Next, for a given $\RR$, we solve the following subproblem to update $\bar{\F}$: \vspace{-3mm}

{
\small
\begin{subequations}\label{mainP:GEE:equiv:updateF}
\begin{align}
\underset{\bar{\F}}{\min} \,
&\sum_{i\in\K_R}\sum_{f_k\in\FF^{*}}\sum_{m\in\MM}
\eta\tau_{f_k}^i\langle{\bar{\E}_{i}}^H\bar{\F}_{f_k,m}\bar{\F}_{f_k,m}^H{\bar{\E}_{i}}\rangle
\\
\mathrm{s.t.}\,
\label{fronthaul:cons}
&\sum_{f_k\in\FF^{*}}\sum_{m\in\MM}\vartheta_{f_k}^i\langle{\bar{\E}_{i}}^H\bar{\F}_{f_k,m}\bar{\F}_{f_k,m}^H{\bar{\E}_{i}}\rangle\leq C_i,\forall i\in\K_R,
\\
\label{rate:cons}
&R_{f_k,m}\leq g_{f_k,m}(\bar{\F}),\forall f_k\in\FF^{*},m\in\MM,,
\\
\label{power:cons}
&\sum_{f_k\in\FF^{*}}\sum_{m\in\MM}\langle{\bar{\E}_{i}}^H\bar{\F}_{f_k,m}\bar{\F}_{f_k,m}^H{\bar{\E}_{i}}\rangle
\leq P_i,\forall i\in\K_R.
\end{align}
\end{subequations}
} \vspace{-3mm}
\begin{figure*}
{
\small
\begin{align}
\label{Gamma}
\nonumber
\Gamma_{f_k,m}^{(n)}(\bar{\F})=
& g_{f_k,m}(\bar{\F}^{(n)})
+ 2\Re
\left\{
\Bigg\langle
\left(\left(\PHI_{f_k,m}^{(n)}-\PI_{f_k,m}^{(n)}(\PI_{f_k,m}^{(n)})^H\right)^{-1}\PI_{f_k,m}^{(n)}\right)^H(\PI_{f_k,m}(\bar{\F}_{f_k,m})-\PI_{f_k,m}^{(n)})
\Bigg\rangle
\right\}
\\
&-
\Bigg\langle
\left(\left(\PHI_{f_k,m}^{(n)}-\PI_{f_k,m}^{(n)}(\PI_{f_k,m}^{(n)})^H\right)^{-1}-(\PHI_{f_k,m}^{(n)})^{-1}\right)^H(\PHI_{f_k,m}(\bar{\F})-\PHI_{f_k,m}^{(n)})
\Bigg\rangle
\end{align}
}\vspace{+2mm}
\hrulefill
\end{figure*}
\begin{algorithm}[!t]
\caption{Joint design of UA, data delivery rate and precoding for F-RANs}
\begin{algorithmic}[1]\label{alg:GEEMax}
{
\small
\STATE \textbf{Initialization}:
Use Alg.~\ref{alg:IFPF} to find a feasible initial point $(\bar{\F}^{(1)}, \{\mu_{k,i},\theta_i\}, \RR^{(1)})$. Set error tolerances $\epsilon_1,\epsilon_2,\epsilon_3>0$. Set $p := 1$.
\REPEAT
\STATE Set $p:=p+1$ and $\kappa:=1$
\REPEAT
\STATE Set $\kappa:=\kappa+1$
\STATE For given $\RR^{(\kappa-1)}$, use $\bar{\F}^{(\kappa-1)}$ as the initial feasible point  and set $n:=1$
\REPEAT
\STATE Update $n:=n+1$
\STATE Find an optimal solution $\bar{\F}^*$ by solving convex problem \eqref{mainP:GEE:equiv:updateF:convexQP}
\STATE Update $\bar{\F}^{(n)}:=\bar{\F}^*$
\UNTIL{$\left|\frac{\PPP_3(\bar{\F}^{(n)})-\PPP_3(\bar{\F}^{(n-1)})}{\PPP_3(\bar{\F^{(n)}})}\right|\leq \epsilon_3$}
\STATE Update $\bar{\F}^{(\kappa)}:=\bar{\F}^*$
\STATE Find an optimal solution $\RR^*$ by solving convex problem \eqref{mainP:GEE:equiv:updateR}
\STATE Update $\RR^{(\kappa)}:=\RR^*$
\UNTIL{$\left|\frac{\PPP_2(\bar{\F}^{(\kappa)},\RR^{(\kappa)})-\PPP_2(\bar{\F}^{(\kappa-1)},\RR^{(\kappa-1)})}{\PPP_2(\bar{\F}^{(\kappa-1)},\RR^{(\kappa-1)})}\right|\leq \epsilon_2$}
\STATE Update $\bar{\F}^{(p)}:=\bar{\F}^*$ and $\RR^{(p)}:=\RR^*$
\STATE Update $\{\mu_{k,i},\theta_i\}$ according to \eqref{mu:update} and \eqref{theta:update}
\UNTIL{$\left|\frac{\PPP_1(\bar{\F}^{(p)},\RR^{(p)})-\PPP_1(\bar{\F}^{(p-1)},\RR^{(p-1)})}{\PPP_1(\bar{\F}^{(p-1)},\RR^{(p-1)})}\right|\leq \epsilon_1$}
}
\end{algorithmic}
\end{algorithm}

\noindent
The key difficulty with subproblem \eqref{mainP:GEE:equiv:updateF} is due to the nonconvex contraint \eqref{rate:cons}. Let $\PHI_{f_k,m}\triangleq\PI_{f_k,m}\PI_{f_k,m}^H+\XI_{f_k,m}$.
Using the first-order Taylor series expansion, we approximate the nonconcave part $g_{f_k,m}(\bar{\F})$ of \eqref{rate:cons} by
its concave lower bound $\Gamma_{f_k,m}^{(n)}(\bar{\F})$ defined in \eqref{Gamma}.
This fact can be proved similarly as in \cite{Tam2016}, and the detailed proof is omitted due to limited space.
The nonconvex subproblem \eqref{mainP:GEE:equiv:updateF} can thus be transformed into the following \emph{convex} quadratic program:
\vspace{+8mm}
\begin{algorithm}[h]
\caption{Finding an initial feasible point for Alg.~\ref{alg:GEEMax}}
\begin{algorithmic}[1]\label{alg:IFPF}
{
\small
\STATE \textbf{Initialization}: Set error tolerance $\epsilon_4>0$.
Find $\bar{\F}^{(1)}$ as $\bar{\F}_{f_k,m}^{(1)}:=\sqrt{\frac{\bar{P}}{\langle\bar{\E}_{7}^H\F_{\text{ran}}\F_{\text{ran}}^H\bar{\E}_{7}\rangle}}\F_{\text{ran}}, \forall f_k\in\FF^{*},m\in\MM$ with a random matrix $\F_{\text{ran}}\in\C^{K_RN_r\times d}$ and $\bar{P} = \frac{P_i}{MK_U}$ until \eqref{fronthaul:cons:intial} and \eqref{power:cons:intial} are satisfied $\forall i\in\K_R$. Set $n := 1$.
\STATE For given $\bar{\F}^{(1)}$, update $\{\mu_{k,i},\theta_{i}\}$ by \eqref{mu:update} and \eqref{theta:update}, and find $\RR$ by solving \eqref{mainP:GEE:equiv:updateR}
\REPEAT
\STATE Update $n:=n+1$
\STATE Find an optimal solution $\bar{\F}^*$ by solving \eqref{maxminP:CQP}
\STATE Update $\bar{\F}^{(n)}:=\bar{\F}^*$
\UNTIL{$\left|\frac{\PPP_4(\bar{\F}^{(n)})-\PPP_4(\bar{\F}^{(n-1)})}{\PPP_4(\bar{\F^{(n)}})}\right|\leq \epsilon_4$}
}
\end{algorithmic}
\end{algorithm}

{
\small
\begin{subequations}\label{mainP:GEE:equiv:updateF:convexQP}
\begin{align}
\underset{\bar{\F}}{\min} \,\,
& \PPP_3(\bar{\F}) \triangleq
 \sum_{i\in\K_R}\sum_{f_k\in\FF^{*}}\sum_{m\in\MM}
\eta\tau_{f_k}^i\langle{\bar{\E}_{i}}^H\bar{\F}_{f_k,m}\bar{\F}_{f_k,m}^H{\bar{\E}_{i}}\rangle
\\
\mathrm{s.t.}\,\,
\label{fronthaul:cons:convexQP}
&\sum_{f_k\in\FF^{*}}\sum_{m\in\MM}\vartheta_{f_k}^i\langle{\bar{\E}_{i}}^H\bar{\F}_{f_k,m}\bar{\F}_{f_k,m}^H{\bar{\E}_{i}}\rangle\leq C_i,\forall i\in\K_R,
\\
\label{rate:cons:convexQP}
&R_{f_k,m}\leq \Gamma_{f_k,m}^{(n)}(\bar{\F}),\forall f_k\in\FF^{*},m\in\MM,
\\
\label{power:cons:convexQP}
&\sum_{f\in\FF^{*}}\sum_{m\in\MM}\langle{\bar{\E}_{i}}^H\bar{\F}_{f_k,m}\bar{\F}_{f_k,m}^H{\bar{\E}_{i}}\rangle
\leq P_i,\forall i\in\K_R.
\end{align}
\end{subequations}
}\vspace{-3mm}


We are now ready to present Alg.~\ref{alg:GEEMax} to solve the original problem  \eqref{mainP:GEE}.
First, to deal with precoding design, the inner loop finds the locally optimal $\bar{\F}$ for a given $\RR$ by iteratively solving the convex problem \eqref{mainP:GEE:equiv:updateF:convexQP}. For the DDR design, the middle loop alternates between solving \eqref{mainP:GEE:equiv:updateF} to find a locally optimal $\bar{\F}$ for a given $\RR$ and solving \eqref{mainP:GEE:equiv:updateR} to find an optimal $\RR$ for a given $\bar{\F}$. Finally, the outer loop updates $\{\mu_{k,i},\theta_i\}$ according to \eqref{mu:update} and \eqref{theta:update} for UA. Alg.~\ref{alg:GEEMax} terminates when there is no improvement in the objective value $\PPP_1(\RR,\bar{\F})$ of \eqref{mainP:GEE}.
\begin{remark}
Let $N_p,N_\kappa$ and $N_n$ respectively be the numbers of iterations of the outer loop, the middle loop and the inner loop when Alg.~\ref{alg:GEEMax} converges. It can be observed that Alg.~\ref{alg:GEEMax} solves problem \eqref{mainP:GEE:equiv:updateF:convexQP} $N_p N_\kappa N_n$ times and problem \eqref{mainP:GEE:equiv:updateR} $N_p N_\kappa$ times.
\end{remark}

%

Alg.~\ref{alg:GEEMax} requires an initial feasible point that satisfies  constraints \eqref{fronthaul:cons}, \eqref{rate:cons} and \eqref{power:cons}.
For this, we consider the following problem: \vspace{-6mm}

{
\small
\begin{subequations}\label{maxminP}
\begin{align}
\underset{\bar{\F}}{\max} \,\,\,
&  \underset{m\in\MM,f_k\in\FF^{*}}{\min} \left\{\frac{g_{f_k,m}(\bar{\F})}{R_{f_k,m}}\right\}
\\
\mathrm{s.t.}\,\,\,
&\sum_{f_k\in\FF^{*}}\sum_{m\in\MM}\vartheta_{f_k,m}^i\langle{\bar{\E}_{i}}^H\bar{\F}_{f_k,m}\bar{\F}_{f_k,m}^H{\bar{\E}_{i}}\rangle\leq C_i,\forall i\in\K_R,
\\
&\sum_{f_k\in\FF^{*}}\sum_{m\in\MM}\langle{\bar{\E}_{i}}^H\bar{\F}_{f_k,m}\bar{\F}_{f_k,m}^H{\bar{\E}_{i}}\rangle
\leq P_i,\forall i\in\K_R.
\end{align}
\end{subequations}
}\vspace{-3mm}

\noindent
Problem \eqref{maxminP} can further be transformed into the following \emph{concave} quadratic program: \vspace{-5mm}

{
\small
\begin{subequations}\label{maxminP:CQP}
\begin{align}
\underset{\bar{\F}}{\max} \,
& \PPP_4(\bar{\F})\triangleq \underset{m\in\MM,f_k\in\FF^{*}}{\min}\left\{\frac{\Gamma_{f_k,m}(\bar{\F})}{R_{f_k,m}}\right\}
\\
\label{fronthaul:cons:intial}
\mathrm{s.t.}\,
&\sum_{f_k\in\FF^{*}}\sum_{m\in\MM}\vartheta_{f_k,m}^i\langle{\bar{\E}_{i}}^H\bar{\F}_{f_k,m}\bar{\F}_{f_k,m}^H{\bar{\E}_{i}}\rangle\leq C_i,\forall i\in\K_R,
\\
\label{power:cons:intial}
&\sum_{f_k\in\FF^{*}}\sum_{m\in\MM}\langle{\bar{\E}_{i}}^H\bar{\F}_{f_k,m}\bar{\F}_{f_k,m}^H{\bar{\E}_{i}}\rangle
\leq P_i,\forall i\in\K_R,
\end{align}
\end{subequations}
}\vspace{-3mm}

\noindent
the solution of which can be found by Alg.~\ref{alg:IFPF}.

\textbf{Convergence Analysis}:  For the inner loop of Alg.~\ref{alg:GEEMax}, the optimal solution $\bar{\F}^{(n)}$ of convex problem \eqref{mainP:GEE:equiv:updateF:convexQP} is feasible to the nonconvex problem \eqref{mainP:GEE:equiv:updateF} and is also better than $\bar{\F}^{(n-1)}$, i.e., {$\PPP_3(\bar{\F}^{(n)})\leq \PPP_3(\bar{\F}^{(n-1)})$}.
Once initialized from a feasible point $\bar{\F}^{(0)}$ by using Alg.~\ref{alg:IFPF}, the inner loop generates a sequence $\{\bar{\F}^{(n)}\}$ of improved feasible solutions for the nonconvex program \eqref{mainP:GEE:equiv:updateF}, and it will eventually converge to a {locally optimal} solution of \eqref{mainP:GEE:equiv:updateF}.
The middle loop utilizes the alternating optimization framework that solves a series of convex problems \eqref{mainP:GEE:equiv:updateR} and \eqref{mainP:GEE:equiv:updateF:convexQP}, and is guaranteed to converge \cite{Tam2016TWC}.
By choosing $c_1=\frac{1}{\ln(1+\tau_1^{-1})}$ and $c_2=\frac{1}{\ln(1+\tau_1^{-2})}$, the outer loop can be proven to be a special case of the majorization-minimization algorithm \cite{Dai2016} and is thus guaranteed to converge.
\vspace{-0mm}

\section{Numerical Results}
\label{Simulation}
\begin{figure}[t!]
  \centering
  \includegraphics[width=0.4\textwidth]{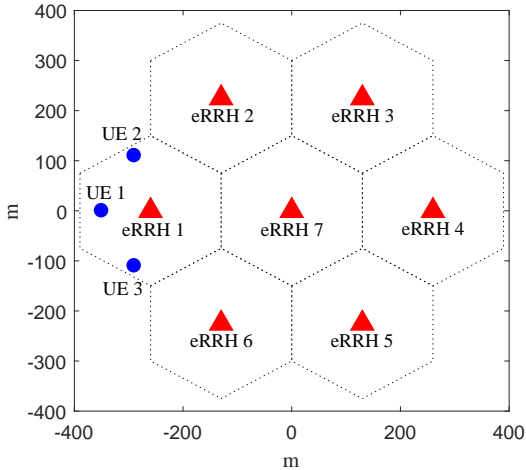}
  \caption{Network scenario used in simulations}
  \label{Fig:2}
\end{figure}
We consider the network scenario in Fig.~\ref{Fig:2} where the locations of the $K_R=7$ eRRHs and $K_U=3$ UEs are fixed.
LTE parameters in Table~\ref{table:parameters} are used in our simulations \cite{3GPP-R9}. Assume that each eRRH is equipped with $N_r=5$ antennas and each UE with $N_u=2$ antennas. We set $P_i=P$, $C_i=C^{FH}$ $\forall i\in\K_R$ and $\SIGMA_k = \sigma^2\pmb{I},\forall k\in\K_U$. At each eRRH, the active mode and the sleep mode consume $84$W and $56$W, respectively \cite{Dai2016}. Constant regulation factors are set as $\tau_1=10^{-5}, \ \tau_2=10^{-3}$. The slope of transmit power is $\beta_i=\beta=2.8$ and $\alpha_i=\alpha=5$ $\forall i\in\K_R$ \cite{Dai2016}.
The error tolerances for the proposed algorithms are taken as $\epsilon_1=10^{-3}$ and $\epsilon_2=\epsilon_3=\epsilon_4=10^{-2}$.
We set $R_{\text{QoS}}=0.1$Mbps,
$\bar{S}=40$Mbps and $M=2$. The numerical result is obtained by averaging over 100 independent channel realizations.

\begin{table}[!h]
\caption{System parameters used in simulations}
\vspace{-2mm}
\label{table:parameters}
\begin{center}
\begin{tabular}{|c|c|}
\hline
Distance between adjacent eRRHs      & $0.3$ km\\
Total bandwidth        & $10$ MHz\\
Std. deviation of log-normal shadowing      & $10$ dB\\
Path loss at distance $d$ (km)
& $140.7+36.7\log_{10}(d)$ dB \\
Noise variance $\sigma_k^2=\sigma^2$
& $-174$ dBm/Hz\\
Maximum eRRH transmit power
& $24$ dBm\\
\hline
\end{tabular}%
\end{center}
\end{table}
\vspace{+0mm}
\begin{table}[!h]
{
\renewcommand{\arraystretch}{1.4}
\tiny
\caption{Example {cache state information} used in simulations}
\vspace{-6mm}
\label{table:cacheplacement}
\begin{center}
\resizebox{0.49\textwidth}{!}{
\begin{tabular}{|c|c|c|c|c|c|c|c|}
\hline
$c_{f_k,m}^i$& $i=1$& $i=2$& $i=3$& $i=4$& $i=5$& $i=6$& $i=7$\\
\hline
$(f_{1},1)$, $(f_{1},2)$    &$1,1$  &$0,0$  &$0,1$  &$0,0$  &$1,0$  &$0,1$  &$0,0$\\
\hline
$(f_{2},1)$, $(f_{2},2)$    &$1,1$  &$0,1$  &$0,0$  &$1,0$  &$0,0$  &$0,1$  &$0,1$ \\
\hline
$(f_{3},1)$, $(f_{3},2)$    &$1,1$  &$0,0$  &$1,0$  &$0,0$  &$1,0$  &$1,0$  &$0,1$\\
\hline
\end{tabular}}
\end{center}
}
\end{table}
In the simulation, each eRRH's caching capacity is set as $B_i=B=\xi SF, \ \forall i\in\K_R$ where $\xi$ indicates the fractional caching capacity. Each eRRH can store a maximum of $\lfloor\xi FM\rfloor$ subfiles randomly chosen from the file library.
Each UE randomly and independently requests one file from a library of $F=6$ files.
{We pick one specific set $\FF^{*}$ of the requested files. The cache state information $\{c_{f_{k},m}^{i}\}$ in \eqref{CDset} are recorded in Table \ref{table:cacheplacement} by checking $\FF^{*}$ against the local caches of eRRHs.
Since we do not consider the caching problem,
we are allowed to selectively place the UEs close to eRRH $1$ where all the requested files are cached.
This arrangement helps us reveal the potential gains resulting from an intelligent caching strategy.


To demonstrate the advantages of the joint design offered by Alg.~\ref{alg:GEEMax}, we evaluate the performance of Alg.~\ref{alg:GEEMax} with local caching (referred to as Alg.~\ref{alg:GEEMax}-C) and Alg.~\ref{alg:GEEMax} without local caching (referred to as  Alg.~\ref{alg:GEEMax}-NC).
We also implement the sole precoder design with local caching (referred to as SPD-C) by modifying Alg.~\ref{alg:GEEMax} such that each UE is always connected to all eRRHs.

{Fig.~\ref{Fig:3}(a) compares the throughput performance among Alg.~\ref{alg:GEEMax}-C, Alg.~\ref{alg:GEEMax}-NC and SPD-C. Attention should be paid to the region of small $\eta$ where problem \eqref{mainP:GEE} prioritizes throughput maximization.
When the fronthaul capacity is limited ($C^{FH}=50$Mbps), the sum rate by Alg.~\ref{alg:GEEMax}-C and Alg.~\ref{alg:GEEMax}-NC are approximately $1.5$ times higher than that by the SPD-C scheme. 
This is because the joint designs show their advantages over the SPD-C in terms of reducing the bottleneck in the fronthaul links [see \eqref{eq:C_FH}] to improve the throughput.
However, for the ample fronthaul capacity of $C^{FH}=1$Gbp, the local caching at the eRRHs or the fronthaul traffic offload via selective UE-eRRH associations offers almost no throughput advantage.
In this case, there is virtually no bottleneck in transferring data traffic from the BBU to each eRRH.
When $\eta=10^{-6}$, the SPD-C even offers higher throughput than that offered by the Alg.~\ref{alg:GEEMax}-C and Alg.~\ref{alg:GEEMax}-NC since it better exploits the coherence combining gain.}

Fig. \ref{Fig:3}(b) demonstrates the power consumption incurred by the three considered schemes. To have meaningful comparisons, we consider the ``busy" power by discounting the fixed eRRH sleeping power from the total power as $P_\text{busy}\triangleq P_{\text{total}}-P_{s}$. From the figure, Alg.~\ref{alg:GEEMax}-C and Alg.~\ref{alg:GEEMax}-NC are the two best performers, where they save approximate $50\%$ of the power consumption for large $\eta$ and more than $66\%$ for small $\eta$ over the SPD-C scheme.
This is because the proposed joint design can effectively save power by: (i) using fewer active eRRHs and less transmission power in the fronthaul links by assigning UEs to appropriate eRRHs, and (ii) reducing the transmission power in the access links by designing effective precoders.
\begin{figure}[t!]
  \centering
  \subfigure[Average total throughput]{\includegraphics[width=0.5\textwidth]{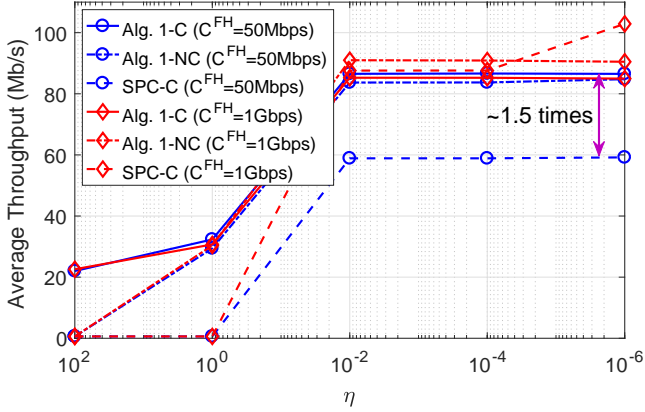}}\\
  \subfigure[Average total busy power]{\includegraphics[width=0.5\textwidth]{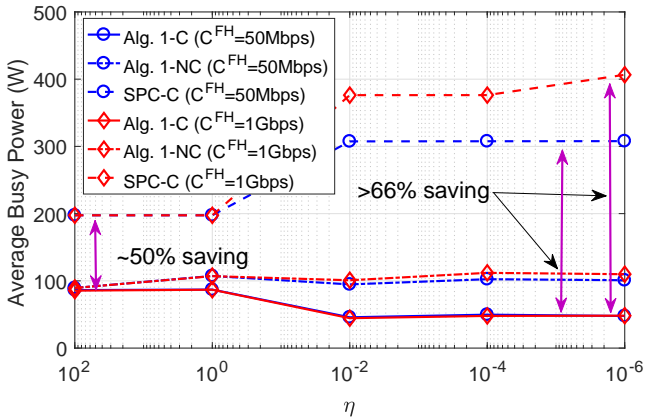}}\\
  \caption{Effects of joint design, local caching and limited backhaul capacity on network performance}
  \label{Fig:3}
\end{figure}

\begin{figure}[t!]
  \centering
  \includegraphics[width=0.5\textwidth]{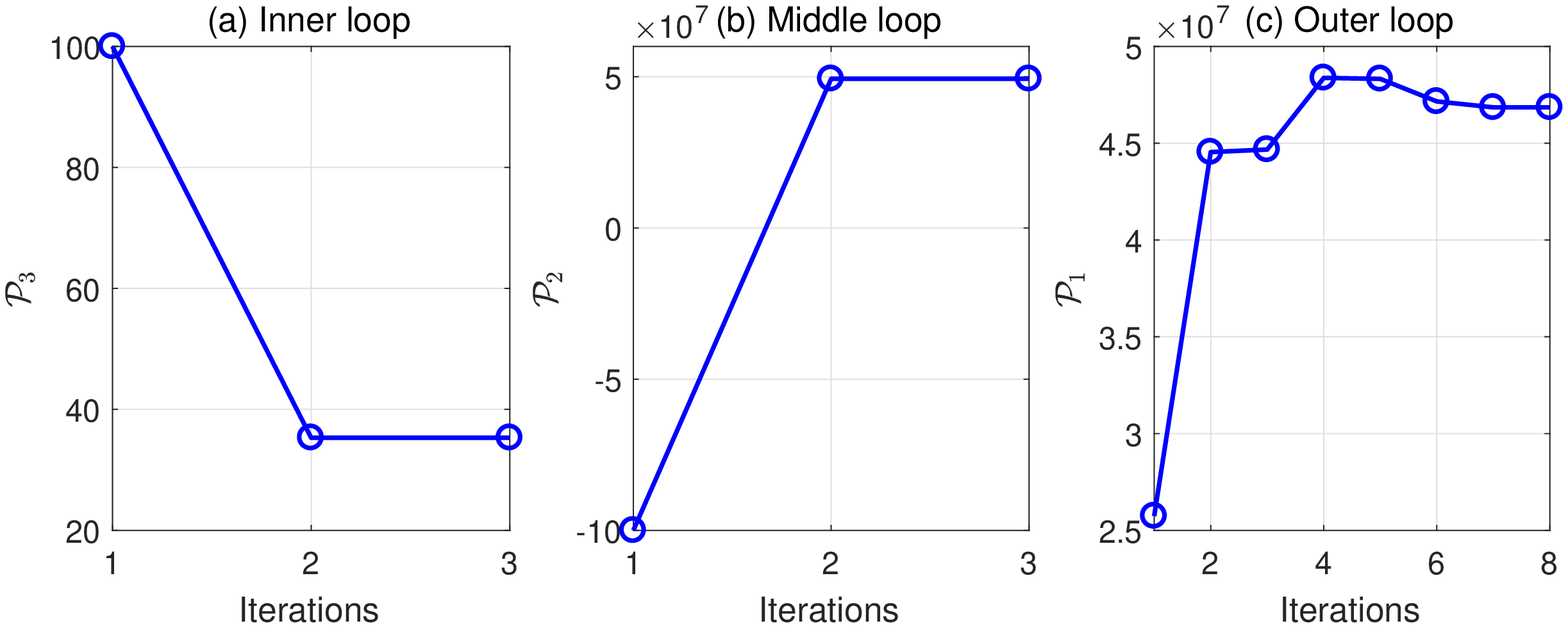}
  \caption{Convergence behavior of Alg.~\ref{alg:GEEMax}}
  \label{Fig:4}
\end{figure}
Fig.~\ref{Fig:4} illustrates the convergence behavior of Alg.~\ref{alg:GEEMax}-C with ${C}^{\text{FH}}=50$Mbps. Figs.~\ref{Fig:4}(a), \ref{Fig:4}(b) and \ref{Fig:4}(c) plot the convergence of the objective functions of the inner, the middle and the outer loops, respectively. In total, the proposed algorithm requires fewer than $50$ iterations to converge, where each iteration corresponds to solving at most two simple convex programs \eqref{mainP:GEE:equiv:updateR} and \eqref{mainP:GEE:equiv:updateF:convexQP}.

\section{Conclusions}
\label{sec:Conclusion}
This paper has studied the joint design of user association, data delivery rate and signal precoding in the downlink of a cache-enabled F-RAN with limited fronthaul capacity. An optimization problem has been formulated with the objective of maximizing the weighted difference of network throughput and total power consumption. The requirements on data delivery rates and maximum eRRH transmit powers have also been included in the design. Applying a range of optimization techniques, we have solved this challenging optimization problem and proposed an iterative algorithm that is guaranteed to converge to obtain a locally optimal solution. Numerical results have shown that our joint design markedly improves both network throughput and power consumption performances of the considered F-RAN.

\section*{Acknowledgment}
This work is supported in part by an ECR-HDR scholarship from The University of Newcastle and in part by the Australian Research Council Discovery Project grant DP170100939.

\bibliographystyle{IEEEtran}
\bibliography{IEEEabrv,newidea2016}
\end{document}